\documentclass{elsart1p}
\usepackage{graphicx}
\usepackage{amssymb}
\usepackage{relsize}
\RequirePackage{xspace}
\def\Babar{\mbox{\slshape B\kern-0.1em{\smaller A}\kern-0.1em B\kern-0.1em{\smaller A\kern-0.2em R}}}
\def\babar{\mbox{\slshape B\kern-0.1em A\kern-0.1em B\kern-0.1em{A\kern-0.2em R}}}

\def\CP{\ensuremath{C\!P}\xspace}

\newcommand{\jprlBase}       {Phys.\ Rev.\ Lett.\xspace}
\newcommand{\jprBase}        {Phys.\ Rev.\xspace}
\newcommand{\jplBase}        {Phys.\ Lett.\xspace}
\newcommand{\nimBaseA}       {Nucl.\ Instrum.\ Methods Phys.\ Res., Sect.\ A\xspace}
\newcommand{\npBase}         {Nucl.\ Phys.\xspace}

\newcommand{\nima}      [1]  {\nimBaseA~{\bf #1}}
\newcommand{\np}        [1]  {\npBase\ {\bf #1}}
\newcommand{\jprl}      [1]  {\jprlBase\ {\bf #1}}
\newcommand{\jprd}      [1]  {\jprBase\ D~{\bf #1}}
\newcommand{\jplb}      [1]  {\jplBase\ B~{\bf #1}}

\def\ccbar {\ensuremath{c\overline c}\xspace}

\def\B       {\ensuremath{B}\xspace}
\def\Bbar    {\kern 0.18em\overline{\kern -0.18em B}{}\xspace}

\def\BB      {\ensuremath{B\Bbar}\xspace}
\def\Bz      {\ensuremath{B^0}\xspace}
\def\Bzb     {\ensuremath{\Bbar^0}\xspace}
\def\BzBzb   {\ensuremath{\Bz {\kern -0.16em \Bzb}}\xspace}
\def\Bu      {\ensuremath{B^+}\xspace}
\def\Bub     {\ensuremath{B^-}\xspace}
\def\Bp      {\ensuremath{\Bu}\xspace}

\def\BpBm    {\ensuremath{\Bu {\kern -0.16em \Bub}}\xspace}

\def\ks{\ensuremath{K^0_S}}

\def\stwob{\ensuremath{\sin\! 2 \beta   }\xspace}

\def\deltat{\ensuremath{{\rm \Delta}t}\xspace}
\def\deltamd{\ensuremath{{\rm \Delta}m_d}\xspace}

\def\to                 {\ensuremath{\rightarrow}\xspace}

\begin{document}

\begin{flushleft}
BABAR-PROC-08/163 \\
\end{flushleft}

\begin{frontmatter}
\title{Update on angles and sides of the CKM unitarity triangle from \babar}
\author{Chih-hsiang Cheng}
\address{California Institute of Technology, Pasadena, California 91125, USA}
\address{For the \babar\ Collaboration}

\begin{abstract}
We report several recent updates from the \babar\ Collaboration on the matrix
elements $|V_{cb}|$, $|V_{ub}|$, and  $|V_{td}|$ of the
Cabibbo-Kobayashi-Maskawa (CKM) quark-mixing matrix, and the angles $\beta$
and $\alpha$ of the unitarity triangle. Most results presented here are using
the full \babar\ $\Upsilon(4S)$ data set.

\end{abstract}
\end{frontmatter}
%
\section{Introduction}
\label{intro}

The Cabibbo-Kobayashi-Maskawa (CKM) quark-mixing matrix~\cite{ref:CKM}
elements $V_{ij}$ represent the couplings of the charged current $W^{\pm}$
between $u_i$ and $d_j$ quarks. In the standard model (SM) of electroweak
interactions, the \CP violation is a consequence of the irreducible phase in
this $3\times 3$ unitary matrix. The main objective of the PEP-II \B-factory
and the \Babar\ detector~\cite{ref:babar_nim} at SLAC is to determine the
parameters in the CKM matrix at high precisions in as many ways as possible. 
If no single set of parameters can satisfy all measurements, one can conclude
that new physics beyond the SM that contributes to \CP violation exists.

The most commonly used unitarity condition of the CKM matrix is 
$V_{ud}V^*_{ub}+V_{cd}V^*_{cb}+V_{td}V^*_{tb}=0$, which forms a triangle on the
complex plane. Its angles and sides represent the more poorly known parameters
of the CKM matrix. In this report, we present some recent updates on the
angles
($\alpha={\rm arg}[-(V_{td}V^*_{tb})/(V_{ud}V^*_{ub})$], 
$\beta={\rm arg}[-(V_{cd}V^*_{cb})/(V_{td}V^*_{tb})$])~\footnote{The third
  angel $\gamma={\rm arg}[-(V_{ud}V^*_{ub})/(V_{cd}V^*_{cb})]$ is not
  discussed here due to limited space.}, 
and the magnitudes $|V_{cb}|$, $|V_{ub}|$, and  $|V_{td}|$.

\section{Angles: $\beta$ and $\alpha$ from time-dependent \CP analyses}
\label{angles}

The angles $\beta$ and $\alpha$ are measured from time-dependent \CP
asymmetries in neutral \B meson decays. 
The asymmetry, arising from
the interference between decays with and without $\Bz$-$\Bzb$ mixing
process, is proportional to 
$S_f\sin(\deltamd\deltat)- C_f\cos(\deltamd\deltat)$, where \deltat is the
difference between the proper decay times of the two \B's in $\Upsilon(4S)$
decays, and \deltamd is the $\Bz$-$\Bzb$ mixing frequency.

Neutral \B decays to \CP eigenstates containing a charmonium and a $K^{(*)0}$
mesons, which are dominated by tree-diagram processes, provide a  direct
measurement of \stwob~\cite{BCP}. To a very good approximation,
$S_f=-\eta_f\stwob$ and $C_f=0$ in the SM ($\eta_f$ is the \CP eigenvalue). We
update the analysis of these decay modes using the full \Babar\ $\Upsilon(4S)$
data set (465 million \BB) and obtain $-\eta_fS_f=0.691\pm 0.029\pm 0.014$ and
$C_f = 0.026\pm 0.020\pm 0.016$~\cite{ccbarK}. 

The charmless \B decays dominated $b\to s$ penguin diagrams provide
opportunities to probe new heavy particles in the loops. We update
$\B\to\eta^\prime K^0$ and $\B\to\omega \ks$~\cite{etapk} analyses, as well as
$\B\to\ks\ks\ks$~\cite{ksksks} using the full \Babar\ $\Upsilon(4S)$
data set. In all cases, $-\eta_fS_f$ is consistent with that from $(\ccbar)K$
final state, indicating that large new physics effects in the $b\to s$
penguin loops are unlikely.

The angle $\alpha$ can be measured by studying modes such as $\B\to \pi\pi,
\rho\rho, \rho\pi$, etc. Because of the sizeable contribution from penguin
diagram with a different weak phase, the
$S_f$ term is modified to $S_f =-\eta_f\sqrt{1-C_f^2}\sin(2\alpha-2\Delta\alpha)$. The
correction $\Delta\alpha$ can be resolved with an isospin
analysis~\cite{isospin}. \Babar\ updates the $\pi\pi$~\cite{pipi}
($S_{\pi^+\pi^-}= -0.68\pm0.10\pm0.03$, $C_{\pi^+\pi^-}= -0.25\pm0.08\pm0.02$)
and $\rho\rho$~\cite{rhorho} analyses. The branching fraction
${\cal B}(\B\to\rho^0\rho^0)=(0.92\pm0.32\pm0.14)\times 10^{-6}$ leads
to $|\Delta\alpha|<17.6^\circ$ at 90\% confidence level. Combining \Babar's
$\B\to \pi\pi, \rho\rho, \rho\pi$ results, we obtain $\alpha=
(81.1^{+17.5}_{-4.9})^\circ$~\cite{ckmfitter}.

\section{$|V_{cb}|$ and $|V_{ub}|$ from semi-leptonic \B decays}
The rates of semileptonic decays of \B mesons to charm and charmless final
states are proportional to $|V_{cb}|^2$ and $|V_{ub}|^2$, respectively.
For $b\to c \ell \nu$ we exploit the partial decay rate as a function of
kinematic variables to determine the decay rate and the form factor at zero
recoil point, at which the form factor suffers from the least theoretical
uncertainty~\cite{isgurwise}. We update the analysis of exclusively
reconstructed $B\to D \ell \nu$ with the second \B fully reconstructed in
hadronic decay modes and obtain $|V_{cb}|= (39.8\pm1.8\pm1.3\pm0.9)\times
10^{-3}$~\cite{Dlnu}, where the last error is due to the theoretical
uncertainty in the form factor. We also perform a global fit using
$\B\to D^{(0,-)}X\ell^+\nu$ decays to simultaneously determine the branching
fractions of $B\to D^{(*)}\ell\nu$, and form factor slopes ($\rho^2_{D}$,
$\rho^2_{D^*}$) in a HEQT-based parameterization. From them we obtain
$|V_{cb}|= (39.9\pm0.8\pm2.2\pm0.9)\times 10^{-3}$ using $D\ell\nu$ and 
$|V_{cb}|= (38.6\pm0.2\pm1.3\pm1.0)\times 10^{-3}$ using
$D^{*}\ell\nu$~\cite{globalclnu}.

For $|V_{ub}|$, we measure branching fractions of $\B\to
\{\pi,\eta,\eta^\prime\}\ell\nu$, with the other \B decays semileptonically,
in three bins of momentum transfer $q^2$. We derive $|V_{ub}|$ using several
theoretical form factor calculations and find the value varies between
$(3.6$--$4.1)\times 10^{-3}$ with total error of about 15\%~\cite{pilnu}. 

The inclusive method suffers from large $b\to c \ell \nu$ background. 
Experimental sensitivity comes from phase space region where $b\to c \ell \nu$
is kinematically suppressed. However, extracting $|V_{ub}|$ from a limited
phase space suffers from theoretical uncertainties from nonperturbative shape
functions. 
We update the inclusive $\B\to X_u\ell\nu$
measurement with the second \B fully reconstructed in hadronic decay
modes. $|V_{ub}|$ is extracted by fitting distributions of $q^2$, mass and
momentum of the $X_u$ system. The results range from $(3.9$--$4.9)\times
10^{-3}$~\cite{ulnu}, depending on models. Experimental uncertainty is
approximate 7\%.

\section{$|V_{td}|$ from radiative penguin processes}
Because top quark is heavier than \B mesons, $V_{td}$ is only accessible
through loops in \B decays, either through \B-$\Bbar$ mixing, or $b\to
d\gamma$ penguin diagrams. We measure the branching fractions of $\B\to
\rho(\omega)\gamma$ using the full \Babar\ data set and find 
${\cal B}(\Bp\to\rho^+\gamma)= (1.20^{+0.42}_{-0.37}\pm0.20)\times 10^{-6}$,
${\cal B}(\Bz\to\rho^0\gamma)= (0.97^{+0.24}_{-0.22}\pm0.06)\times 10^{-6}$,
and
${\cal B}(\Bz\to\omega\gamma)= (0.50^{+0.27}_{-0.23}\pm0.09)\times
10^{-6}$~\cite{rhogamma}. 
The ratio $|V_{td}|/|V_{ts}|$ can be extracted via 
$\frac{{\cal B}(\B\to\rho(\omega)\gamma)}{{\cal B}(\B\to K^*\gamma)} =
S|V_{td}/V_{ts}|^2 (\frac{1-m^2_{\rho(\omega)}/m_B^2}{1-m^2_{K^*}/m_B^2})
\zeta^2_{\rho(\omega)} \times [1+\Delta R_{\rho(\omega)}]$. We find
$|V_{td}|/|V_{ts}|= 0.233^{+0.025+0.022}_{-0.024-0.021}$. This value is
consistent with the result from $B_s$ mixing
($0.208\pm0.002^{+0.008}_{-0.006}$)~\cite{PDG} from CDF and D0 
Collaborations. Even though it is less precise than the results from \B
mixing, radiative penguin channels provide a nice independent confirmation.

\section{Conclusions}
\Babar\ ended its $\Upsilon(4S)$ program in late 2007. Within a year, \Babar\
has updated many measurements related to the angles and sides of the CKM
unitarity triangle. There are no significant contradictions among the
measurements under the CKM description of the Standard Model.

\end{document}